\documentclass[11pt,letterpaper]{llncs}
\usepackage{amsmath}
\usepackage{amssymb}
\usepackage{graphicx}
\usepackage{marvosym}
\usepackage{url}
\usepackage{fancyhdr}

\usepackage{geometry}
\newcommand{\keywords}[1]{\par\addvspace\baselineskip
\noindent\keywordname\enspace\ignorespaces#1}

\fancypagestyle{firstpage}{\fancyhf{}
\setcounter{page}{1}
\fancyhead[C]{\small{}}
\fancyfoot[L]{}
\rfoot{\thepage}
}

\begin{document}


\title{\LARGE{A Tree-Based Repository Blockchain Framework for Shared Governance in Collaborative Fork Ecosystems}}


%
%
\author{\large{Razwan Ahmed Tanvir$^1$ and Greg Speegle$^2$}}

\institute{\large{
$^1$Department of Computer Science, Baylor University, Waco, Texas, USA \\
\email{Razwan\_Tanvir1@alumni.baylor.edu} \\
\vspace{0.3cm}
$^2$Department of Computer Science, Baylor University, Waco, Texas, USA \\
\email{Greg\_Speegle@baylor.edu}
}}

%


%
%


\maketitle

\thispagestyle{firstpage}

\begin{abstract}
Collaborative blockchain ecosystems allow diverse groups to cooperate on tasks while providing properties such as decentralization and transaction security. We provide a model that uses a repository blockchain to manage hard forks within a collaborative system such that a single process (assuming that it has knowledge of the requirements of each fork) can access all of the blocks within the system. The repository blockchain replaces the need for Inter Blockchain Communication (IBC) within the ecosystem by navigating the networks. The resulting construction resembles a tree instead of a chain. A proof-of-concept implementation performs a depth-first search on the new structure.
\end{abstract}


\section{Introduction}

Blockchain is a distributed ledger that enables secure, transparent, and tamper-proof transactions. It is the core concept behind Bitcoin and other cryptocurrencies, but it has the potential to be used for a wide variety of other applications, including supply chain management \cite{141}, voting systems, health care, and digital identity.

Blockchain technology operates by establishing a Peer-to-Peer \cite{177} network of computers, each of which keeps a duplicate of the ledger. This ledger is constantly updated with new transactions that are verified by network nodes before being added to the ledger in blocks. Once a block has been added to the ledger, it cannot be changed without the agreement of the majority of network nodes.

Blockchain networks, such as Bitcoin \cite{103} and Ethereum \cite{142}, register transactions through mining. Miners are network nodes that verify and validate transactions before adding them to the blockchain as blocks. Every transaction must adhere to a set of rules that govern transaction creation and processing.

All network participants must follow the protocol's rules. The protocol cannot be updated, altered, or deleted unless the majority of network participants agree. However, there is occasionally disagreement among network participants about whether or not to change the protocol.

If the network accepts a proposed protocol change as a whole, but some participants do not agree with the updated network's terms, they may choose to fork the current network. Forking generates a new network with the modified rules, retaining the old transactions while introducing updated regulations for new transactions and protocol axioms.  

This article aims to foster a unified blockchain ecosystem by storing fork events in a repository blockchain for more diverse blockchain governance \cite{135} and transparent management of interoperable blockchains. A repository blockchain that stores information about forks can be utilized in various ways, providing valuable functionalities for organizations in the blockchain ecosystem. Below are some of the areas where this unified blockchain can increase the organizational benefit for the better.

\subsection{Decentralized Governance Bodies}

The blockchain can be structured to support decentralized governance bodies within the organization, where each fork represents an independent governing entity. This creates a dynamic ecosystem where the root blockchain acts as the parent organization, and its forks function as autonomous sub-governing bodies \cite{159}. Each fork can have its governance structure, enabling adaptability and specialization while maintaining an organizational ordinance.

\subsection{Blockchain Governance and Transparency}

The repository blockchain can serve as a transparent ledger documenting the history of blockchain forks within an organization. This fosters governance transparency \cite{160}, allowing stakeholders to track and understand the evolution of the blockchain \cite{161} network. It aids in decision-making processes related to protocol upgrades, ensuring a clear and auditable history.

\subsection{Fork History Analysis}
The blockchain ecosystem enables organizations to analyze the history of forks \cite{162}, their causes, and their effects on the network. This analysis can help organizations understand the reasons behind forks, evaluate the success or challenges posed by each fork, and inform strategic decisions for the future. It contributes to a more robust and resilient blockchain infrastructure.

\subsection{Collaborative Ecosystem Development}
Organizations can use the blockchain to foster collaboration within the ecosystem, encouraging the development of independent projects and innovations through forks. By allowing forks to operate semi-autonomously, organizations can stimulate creativity and diverse contributions. This collaborative environment promotes a healthy and innovative blockchain ecosystem where different branches can experiment with new features or functionalities.

\subsection{Contribution}
This paper makes the following contributions:
\begin{itemize}
    \item The implementation of a collaborative blockchain environment into a tree-like structure
    \item A depth first search program to process all of the nodes in a heavily forked blockchain environment
\end{itemize}

\section{Background}
This section covers the basic concepts of decentralized ledgers and how they evolve. We examine blockchain components and the nature of fork events. This provides the context needed to understand how our repository model unifies different network histories.
\subsection{What is Blockchain? }
Blockchain is a decentralized digital ledger that facilitates secure and transparent transactions among users in a network. Data redundancy and consistency \cite{100} are ensured in this system by having an identical copy of the ledger kept on file by each network node. Using cryptographic techniques, transactions are grouped into blocks and added to the current chain. This decentralized method has substituted for a central authority, protecting against illegal data modifications. The users of the network must agree to the validity of transactions which makes the ledger transparent to the participants of the blockchain.

The power of blockchain technology comes from a few essential components \cite{146}. First off, data integrity and reliability are strengthened by the use of a distributed ledger, which provides data replication among all network users. Data immutability is ensured by the cryptographic linking of blocks; if the contents of one block are altered, all following blocks must also be altered. Blockchain transactions are irreversible, it takes fresh transactions to undo previous ones. Decentralization strengthens the network by enabling decisions to be taken by consensus among members, which improves resilience. Furthermore, by using public-private key pairings to ensure anonymity, blockchain technology ensures privacy without revealing personal information. Together, these characteristics strengthen blockchain's effectiveness.

 Fundamentally, blockchain uses cryptographic hash functions to produce distinct digital fingerprints that guarantee the security and integrity of the data. These digital fingerprints, also known as hashes, are produced for every block in the blockchain that contains transactional data. 
\begin{figure}[!htb]
    \centering 
    \includegraphics[width=0.5\linewidth]{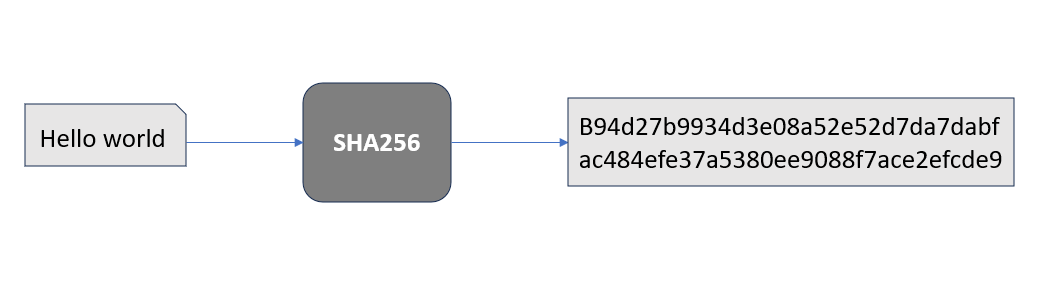}
    \caption{SHA256 hash of the string `Hello world'}
    \label{fig:sha256}
\end{figure}

\begin{figure}[!htb]
    \centering
    \includegraphics[width=0.5\linewidth]{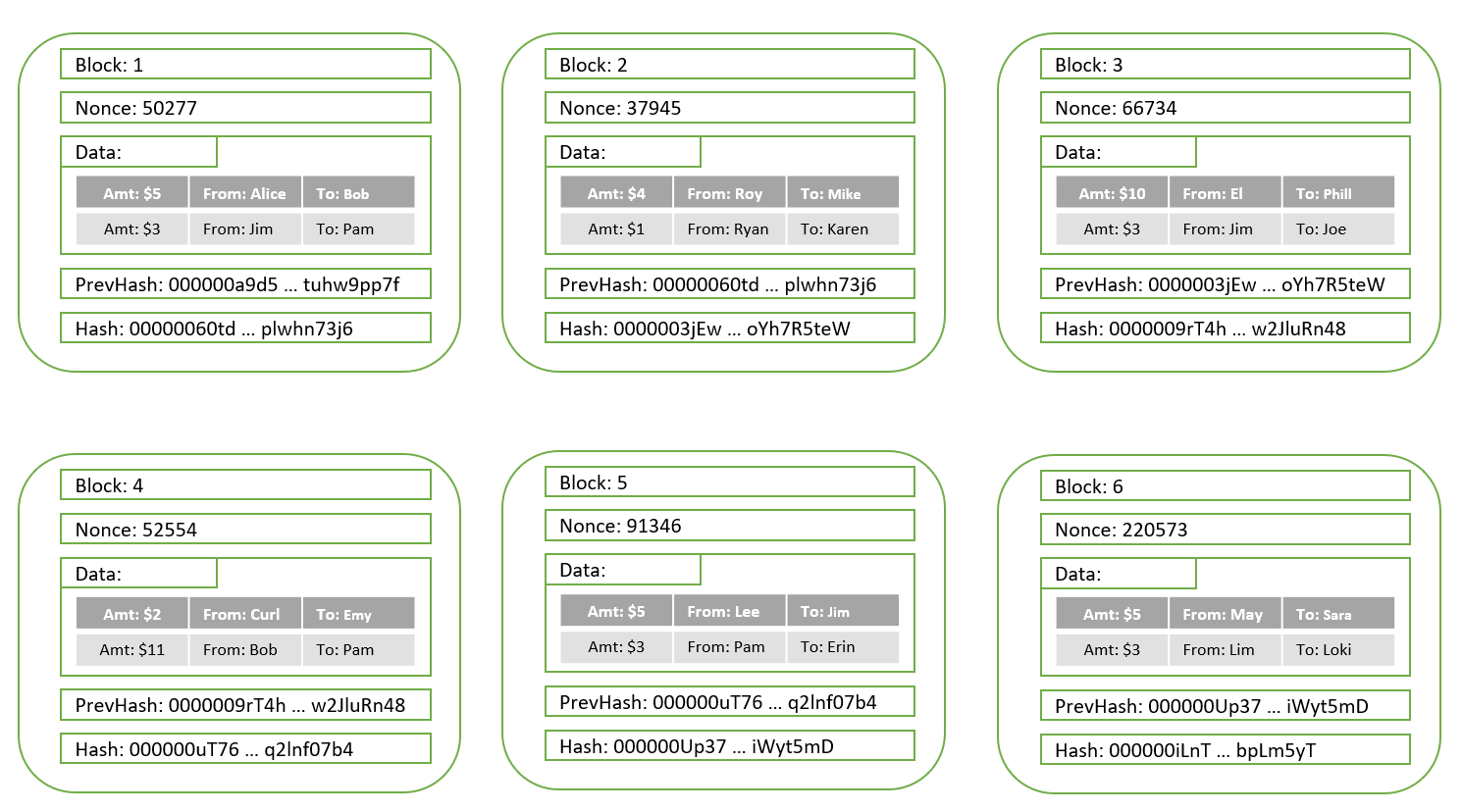}
    \caption{Blockchain of six mined blocks}
    \label{fig:chainblocks}
\end{figure}
The hash of any block will be generated as in Figure \ref {fig:sha256} but the entire contents of the block will be hashed. 
Every block contains the hash of the one before it, creating an unchangeable chain in Figure \ref{fig:chainblocks}. Thus, changing any previous block's data makes the hash of every subsequent block invalid. From this, any mutation in any block can be identified. The question might arise what happens when someone generates the hashes of the subsequent blocks too? The mining of a block with proof-of-work addresses this. There is a constraint that the hash generated for a block must be lower than a numerical target value. This constraint restricts the quick generation of block hashes. 

\begin{figure}[!htb]
    \centering 
    \includegraphics[width=0.5\linewidth]{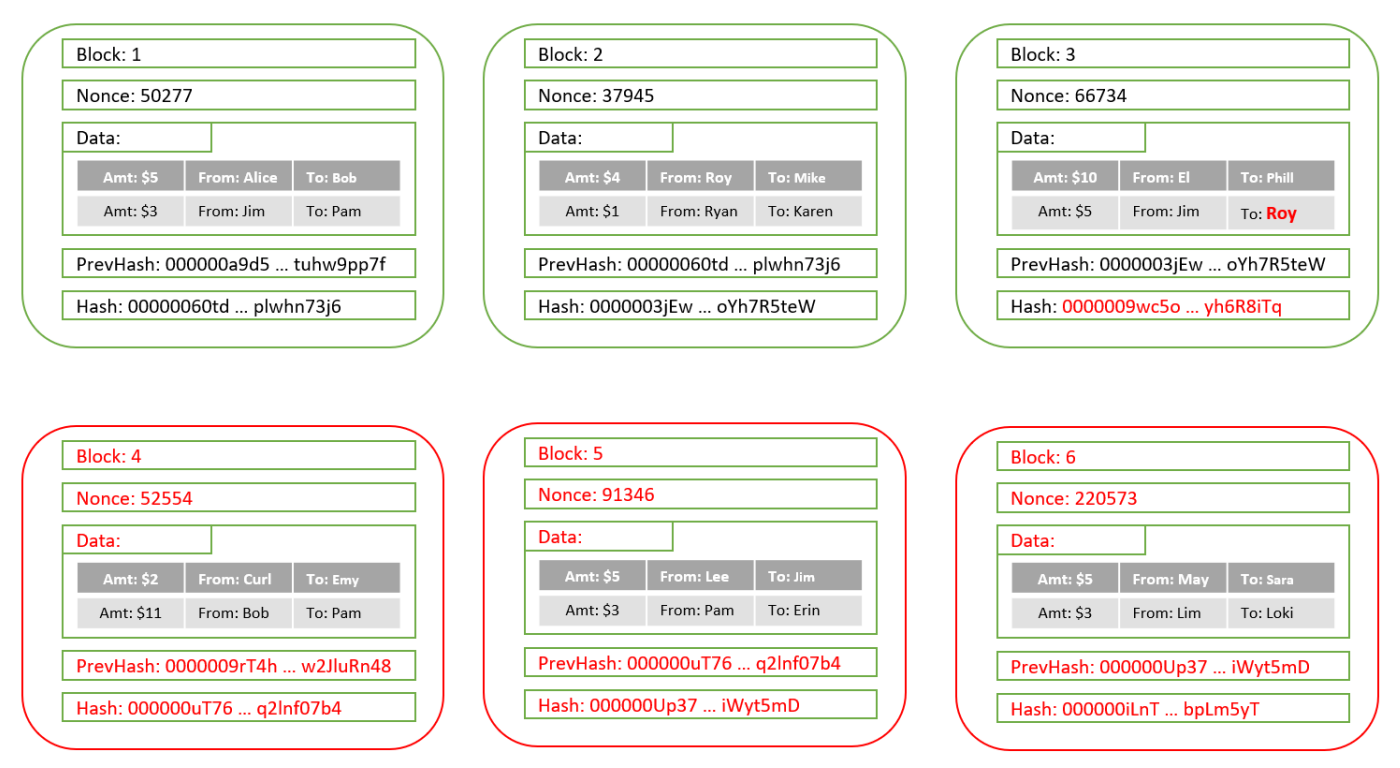}
\caption{Altering block data}\label{fig:altchainblocks}
\end{figure}

Miners in a blockchain validate transactions and secure the network. They do so by repeatedly altering a variable called the `nonce' in a block to generate a unique hash using the SHA-256 cryptographic function. Their goal is to find a hash that meets the network's current difficulty criteria—a hash that is below a certain target value. If successful, the miner can add a new block to the blockchain and is rewarded with newly minted cryptocurrency and transaction fees. Since this search for the hash takes a lot of computational power, altering any previous node requires an enormous amount of energy which ensures the immutability of the blockchain. This energy consumption is the proof-of-work for a transaction added to the blockchain. Moreover, the hacker also needs to change the copy of the ledger in every other node in the network (Figure \ref{fig:altchainblocks}), so it is virtually impossible to change the data of a blockchain network. Consensus procedures and the distributed nature of the ledger guarantee that numerous copies of the blockchain exist on various computers (peers), making it simple to identify discrepancies.

\subsection{Blockchain Forks}
Within the field of blockchains, `forks' represent significant events \cite{130} where changes are made to the rules or software governing the blockchain, resulting in the splitting of the blockchain's history into distinct directions. Forks enable the execution of these modifications in the blockchain. However, it's essential to clarify that forks themselves do not achieve consensus but rather serve as a mechanism to propose and implement changes. The occurrence of a blockchain fork might result in the creation of two distinct blockchains, each possessing its distinct features and governing rules. This has the potential to create novel prospects and enhance the diverse nature of blockchain applications.

\subsection{Types of Forks }

The blockchain forks can be categorized into two primary classifications: hard and soft. Hard and soft forks \cite{131} are fundamental concepts providing distinct methods for incorporating modifications to a blockchain's governing principles. A hard fork refers to a substantial and unresolvable modification made to the protocol of a blockchain, leading to a distinct and permanent separation in the transaction history of the blockchain. The implementation of this modification results in the establishment of rules that are incompatible with the preceding version of the blockchain's software. A hard fork results in the formation of a distinct and independent blockchain.

On the other hand, a soft fork denotes a comparatively subtle modification to the regulations of a blockchain that maintains compatibility with previous versions. In the event of a soft fork, the newly implemented rules are specifically formulated to be compatible with the previous software version, hence guaranteeing the absence of instantaneous incompatibility. Unlike a hard fork, which necessitates that all nodes upgrade and accept the new version, this type of fork just needs the majority of miners to upgrade to enforce the new rules. An example of a soft fork in the context of Bitcoin is the introduction of Segregated Witness \cite{158} (SegWit). The modification altered how data was saved within the Bitcoin blocks while maintaining compatibility with the previous type of transactions. Soft forks facilitate a more seamless transition by enabling the coexistence of old and new software, hence preventing a total splitting in the transaction history of the blockchain. Both hard forks and soft forks play essential roles in the governance and evolution of blockchain networks.

Forks are an inherent and recurrent phenomenon within the framework of any blockchain system. While existing research predominantly views forks as mechanisms for mitigating disparities among network participants and engendering new cryptocurrencies, this study endeavors to make a distinctive contribution. Specifically, the primary objective of this study is to foster a unified blockchain ecosystem by preserving fork history in a repository blockchain.

\section{Related Work}

Research and innovation on collaborative blockchain \cite{163} ecosystems, interoperability, and governance are widespread. Scholars have studied blockchain technology to improve decentralized network communication \cite{132} and interoperability \cite{133} in cross-blockchain \cite{173} scenarios. Previous publications recognized the importance of hard forks in creating new blockchains \cite{164}, providing the groundwork for talks on improved communication and compatible ecosystems. However, the existing research provides little attention to unifying the blockchains generated from hard forks and undermines the diverse applications of a blockchain ecosystem. This project presents the concept of a `Repository Blockchain', a unified framework for integrating blockchains that originate from hard forks of a primary blockchain, along with their subsequent forks. The paper delivers a proof-of-concept for this ecosystem, showcasing its ability to facilitate tree traversal in a collaborative blockchain environment. Moreover, integrating smart contracts within the repository blockchain makes the ecosystem autonomous, robust, and trustless \cite{175}.



\subsection{Inter-Blockchain communication}
The increasing development of the blockchain industry has led to a frequent emergence of hard forks, resulting in the creation of multiple interconnected blockchains. The decentralized expansion presents a significant difficulty, namely, the need to develop seamless communication among the various separate blockchains. The notion of IBC is introduced as a crucial solution to tackle the difficulty of enabling the transfer of information and assets between different blockchain networks \cite{134}. Academics showed interest in IBC due to its potential to effectively facilitate a unified and integrated blockchain ecosystem. The smooth transmission of value and data between diverse blockchains is of utmost importance to fully harness the potential of decentralized applications and establish consistent interoperability within the blockchain domain. In this work, hard forks are integrated into the repository blockchain, and the communication between the forks is established through smart contracts.

\subsection{Blockchain Governance}
Blockchain governance \cite{135} is a key component within the dynamic blockchain environment, involving the systems and procedures via which decisions are formulated and executed inside a decentralized network. The significance of blockchain governance becomes increasingly crucial as the number of interconnected blockchains expands through hard forks and other expansion techniques. IBC plays an important role in the establishment of a robust blockchain governance ecosystem. 

The academic research about blockchain governance encompasses a wide range of topics, which resemble the complex and varied challenges and prospects inherent in decentralized systems. A prominent area of interest lies in the examination of Decentralised Autonomous Organisations (DAOs) \cite{143}, wherein scholars explore the complex processes of constructing and executing smart contract-driven organizations that possess decentralized mechanisms for decision-making. Simultaneously, scholarly inquiries into the economic dimensions of blockchain governance, including the study of tokenomics \cite{144}, are influencing conversations around the impact of economic incentives on decision-making processes inside network structures. In addition to technical factors, scholars are currently addressing the security concerns \cite{136} of governance methods and their compatibility with legal frameworks.



\subsection{Blockchain ecosystem}
The forking of a blockchain often gives rise to a dynamic and multifaceted ecosystem \cite{166}, characterized by innovation, competition, and community engagement. When a blockchain undergoes a fork, whether it be a hard fork resulting in the creation of a new, separate blockchain \cite{167}, or a soft fork leading to a temporary divergence in protocol, it sets the stage for the evolution of a diverse ecosystem.

At the heart of this ecosystem are the different chains themselves, each with its own set of rules, protocols, and community of participants. These chains may compete for dominance, each vying to attract users, developers, and miners to their network by offering unique features, improved scalability, or enhanced security.

Within this competitive landscape \cite{168}, innovation thrives as developers on each chain work to improve and differentiate their platforms. New technologies, applications, and protocols emerge, contributing to the overall growth and evolution of the blockchain ecosystem.

Interoperability between divergent chains also becomes increasingly important as the ecosystem matures. Solutions such as cross-chain communication protocols \cite{169}  and repository blockchains enable seamless interaction and value transfer between different blockchain networks, improving collaboration and expanding the scope of possibilities within the ecosystem. 

\section{Methodology}
This section outlines the practical implementation of the repository blockchain and its supporting infrastructure. We describe the deployment of multiple test networks and the smart contract logic used to manage fork metadata. These steps demonstrate how a theoretical tree structure is realized in a decentralized environment.
\subsection{Repository Blockchain} 
In the context of establishing a collaborative blockchain ecosystem involving several autonomous entities, the development of a repository blockchain emerges as a solution. The repository blockchain operates as a hub for essential data regarding several interconnected networks within the ecosystem. A novel methodology for creating the repository blockchain involves the inclusion of fork events, which occur when the root blockchain is duplicated (hard fork), resulting in the creation of a new blockchain. This establishes an archive of fork history, which documents relevant information of each blockchain instance.

This repository serves as the foundational structure of the ecosystem, comprising of data elements including the network ID, parent network ID, fork ID, timestamp, and comprehensive particulars associated with each individual fork. In essence, it creates a graphical representation of a Blockchain Fork Tree, illustrating the progression of the root chain over the course of time. Every instance of a fork event is recorded which allows traversal within the ecosystem itself. A repository blockchain is structured similarly to a traditional Blockchain but records metadata related to blockchain forks. Each block within the repository blockchain contains information about a specific fork event such as- 
parent blockchain network ID, child blockchain network ID, date and time of the fork, details of the governance decisions leading to the fork, and technical specifications of the forked network.

This repository blockchain allows the traversal of the ecosystem to better understand the evolution of blockchain networks over time. For the experimental observation, we ignored the descriptive details of the fork events such as DateTime, specifications, and particulars at the time of implementation. 

The advantages of this repository blockchain reside in its ability to streamline communication, promote interoperability, and facilitate decision-making processes across diverse networks. The repository might be utilized as a dynamic point of reference, enabling the seamless navigation of participants within the unified blockchain ecosystem. Additionally, this fork archive could serve to enhance transparency and governance within the broader system, providing stakeholders with a thorough comprehension of the ecosystem.

\subsection{Deploying Ethereum Test Networks}
Several Ethereum test networks were deployed to demonstrate that the proposed system can store the fork history evolving from a root chain. For the experiments, seven such test networks were deployed, in which, one is the repository blockchain, and the other six blockchains form the blockchain ecosystem. Each network is uniquely identified by their network ID which is given in the genesis file of the deployed blockchain.

The proposed system uses a Geth (1.11.4-stable) client to deploy test networks in the Windows environment. Go Ethereum (Geth) is a command-line interface for Ethereum development since it provides many essential functionalities. It can support Ethereum mining, verify and broadcast blocks and transactions, and operate an entire Ethereum node. Geth is also used to develop and test Ethereum applications, deploy and interact with smart contracts, and manage large transaction volumes. It is a reliable and secure program that has been used for running Ethereum nodes for a long time. Its extensive documentation and large user and developer community further support its utility, making it an excellent choice for reducing application development overhead.



\subsection{Smart Contract for the Repository Blockchain}
A smart contract was created to store fork history in the repository blockchain. The smart contract was written in Solidity, which is the programming language used to develop Ethereum smart contracts. Remix online IDE was used to compile and test the smart contract before the deployment. The Remix IDE generated the contract \textit{.abi} and \textit{.bin} file. The smart contract provides the following functionalities: 
    \begin{enumerate}
        \item \textbf{addForkDetail(networkId, portNumber, 
        parentId, forkBlockNo)}: This function allows users to add a new fork to the repository blockchain. 

        \item \textbf{findForkId(networkId)}: This function allows users to find the fork ID for a given network ID for event inspection. The function takes the network ID as an argument and returns the fork ID, or type(uint256).max if the fork is not found.
        
        \item \textbf{getForkData(forkId)}: This function allows users to get the data for a specific fork. 
        
        \item \textbf{getChildren(forkId)}: This function allows users to get the child forks of a given fork. The function takes the fork ID as an argument and returns an array of the fork's child fork IDs.

        \item \textbf{getAllForkDetails()}: This function returns all the fork events to the caller of the function from web3.js. The adjacency list forming the parent-child relationship among the network IDs is prepared from the fork details.

    \end{enumerate}

Steps on how to run a private blockchain and deploy smart contracts on blockchain network can be found on the github repository mentioned in the result section.

\section{Results}
This section describes the details on how the blockchain ecosystem can be traversed using the Depth First Search (DFS) tree traversing algorithm. The implementation of the ecosystem in \texttt{node.js} including the process of storing the fork events, adding transactions in blockchains, and searching for data in the ecosystem can be found {here\footnote{https://github.com/RazwanTanvir/BlockchainForkTree}}. This proof-of-concept further entails that the class of other tree algorithms such as Level Order Traversal or Breadth First Search(BFS) can also be implemented on the ecosystem.

\subsection{Blockchain Structure}

Let $G = (V, E)$ represent the blockchain ecosystem, where:
\begin{itemize}
  \item $V$ is the set of interconnected blockchains, each denoted by a unique network identifier $n_i$.
  \item $E$ is the set of edges representing forks between blockchains. Each edge $(n_i, n_j)$ indicates that blockchain $n_i$ is forked to form blockchain $n_j$. 
\end{itemize}

\subsection{Adjacency List}

The adjacency list $\mathcal{A}$ is a mapping from each blockchain $n_i$ to the list of its forked blockchains. Mathematically, it can be represented as:
\[ \mathcal{A}(n_i) = \{ n_j \ | \ (n_i, n_j) \in E \} \]

\subsection{Depth-First Search (DFS)}

The Depth-First Search algorithm is used to traverse the blockchain ecosystem in search of specific data. It starts from a root blockchain and explores as far as possible along each branch before backtracking.

The DFS algorithm is defined recursively as follows:
\[
\text{DFS}(n_i) = 
\begin{cases} 
\text{Found}, & \text{if target data is in } n_i \\
\text{DFS}(n_j), & \text{for each neighbor } n_j \text{ of } n_i \\
\text{Not Found}, & \text{if data not found in any neighbor}
\end{cases}
\]

\subsection{Searching Data in the Ecosystem}
\begin{figure}[t!]
 \centering
 \includegraphics[width=0.5\textwidth]{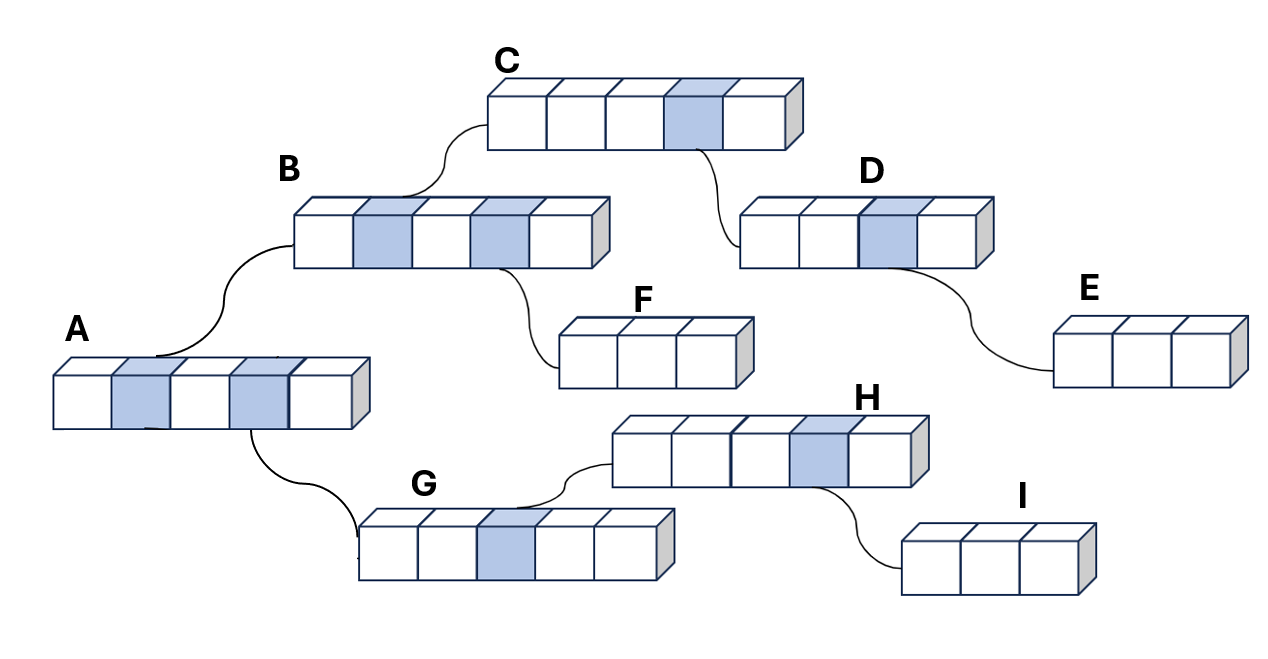}
 \caption{Blockchain Fork Tree (Blue blocks indicate fork events)}
 \label{fig:forktree}
\end{figure}

To search for a specific value within the Figure \ref{fig:forktree} blockchain ecosystem, the DFS algorithm is applied as follows:

\begin{enumerate}
    \item Begin by reading the repository blockchain which contains a record of all fork events.
    \item Construct an adjacency list based on the repository blockchain's data to map the relationships between a blockchain and its forks. For this scenario, an adjacency list like A $\rightarrow$ B, G; B $\rightarrow$ C, F; D $\rightarrow$ E; G $\rightarrow$ H; H $\rightarrow$ I signifies that blockchain ``A'' has forks ``B'' and ``G'', ``B'' further forks into ``C'' and ``F'', and so on.
    \item Start the DFS traversal from blockchain ``A''. Check for the presence of the target data within ``A''. If the target data is found within ``A'', return the block containing the value and terminate the search process. If the target data is not found in ``A'', recursively initiate DFS for each of ``A's'' forks, following the adjacency list. The search would first explore ``B'' then ``G'' as they are direct forks of ``A''.
    \item For each fork encountered (e.g., ``B'' and ``G''), repeat the DFS process: check the current blockchain for the target data, and if not found, recursively search its forks as per the adjacency list. After exploring ``B'' to its depth (through ``C'', ``D'', ``E''), backtrack and then explore ``F'' before exploring ``G'', followed by ``H'', and finally ``I''.
    \item The search process continues until either the target data is found or all blocks in the blockchain ecosystem have been traversed without finding the target data.
\end{enumerate}

The figure \ref{fig:forktree} illustrates a blockchain structure with a main chain and multiple forks, represented as blocks.

\section{Experimental Constraints and System Limitations}

While the proof-of-concept successfully utilizes a Depth-First Search (DFS) algorithm to traverse the interconnected network architecture, the experimental setup possesses distinct limitations that contextualize the current findings. 

\subsection{Localized Deployment Environment}
The experimental blockchain ecosystem was simulated using only seven Ethereum test networks deployed via a Geth (1.11.4-stable) client on a single Windows host machine. 
While Geth is highly efficient and capable of running full Ethereum nodes, this localized environment does not replicate the network latency or the complex, decentralized competitive landscape found in a live, global blockchain ecosystem.

\subsection{Intentional Metadata Omission}
To facilitate the experimental observation and simplify the proof-of-concept, critical descriptive details regarding the fork events were intentionally ignored. 
Omitted data included precise DateTime stamps, technical specifications, and specific implementation particulars present at the time of the forks.

\subsection{Absence of Performance Metrics}
The current study serves primarily as a functional proof-of-concept for decentralized data management and retrieval. 
Consequently, comprehensive quantitative evaluations regarding the system's scalability, computational efficiency, and performance under realistic transaction volumes remain unaddressed, necessitating formal performance evaluation in future research.

 \section{Conclusion}
The proof-of-concept demonstrates the potential of repository blockchain systems in efficiently searching and accessing data across interconnected networks. By employing depth-first search algorithm and smart contract queries, the system effectively navigates through blockchain ecosystems to locate specific data points. This showcases the feasibility of leveraging repository blockchains for decentralized data management and retrieval. Additionally, the use of standardized information storage and retrieval mechanisms enhances interoperability, while advancements in governance frameworks promote transparency in decision-making processes. 

Moving forward, future research can focus on enhancing scalability, efficiency and security through advanced smart contract tailored to repository blockchains. Attention to the performance evaluation of these frameworks are due. Moreover, exploring novel applications in sectors like healthcare, finance, and supply chain management offers promising avenues for sector-specific solutions where a transparent environment is necessary in a shared network. 

\bibliographystyle{unsrt}
\bibliography{references}

\vspace{2cm}

\section*{Authors}
\noindent {\bf Razwan Ahmed Tanvir} \\Razwan's primary academic research focuses on the advancement of blockchain technology, specifically exploring Inter-Blockchain Communication (IBC), smart contracts, and decentralized governance within collaborative ecosystems. In addition to his current work on developing framework solutions using repository blockchains, he has previously contributed to published research regarding the application of blockchain and IPFS technologies to secure scientific paper peer-reviewing systems.\\

\noindent {\bf Greg Speegle} \\
Dr. Speegle was an assistant professor at Baylor University from 1990-1996; an associate professor from 1996-2006; a professor from 2006 on. He served as chair of the computer science department from 2011-2018. 
He works on two distinct projects. The first is CT³ : Central Texas Computational Thinking, Coding and Tinkering, which is designed to train certified public school teachers to become certified computer science teachers. The second is development of a general parallelization theory in order to advance automated parallel computation beyond MapReduce. \\

\end{document}